# 100% Reliable Frequency-Resolved Optical Gating Pulse-Retrieval Algorithmic Approach

Rana Jafari*, and Rick Trebino
*Georgia Institute of Technology, School of Physics, Atlanta, GA 30332 USA*
*rjafari7@ gatech.edu

***Abstract*—Frequency-resolved optical gating (FROG) is widely used to measure ultrashort laser pulses, also providing an excellent indication of pulse-shape instabilities by disagreement between measured and retrieved FROG traces. However, FROG requires—but currently lacks—an extremely reliable pulse-retrieval algorithm. So, this work provides one. It uses a simple procedure for *directly* retrieving the *precise pulse spectrum* from the measured trace. Additionally, it implements a multi-grid scheme, also quickly yielding a vastly improved guess for the spectral phase before implementing the entire measured trace. As a result, it achieves 100% convergence for the three most common variants of FROG for pulses with time-bandwidth products as large as 100, even with traces contaminated with noise. Here we consider the polarization-gate (PG) and transient-grating (TG) variants of FROG, which measure amplified, UV, and broadly tunable pulses. Convergence occurs for all of the >20,000 simulated noisy PG/TG FROG traces considered and is also faster.**

## I. INTRODUCTION

FREQUENCY-resolved optical gating (FROG), introduced in 1991, solved the long-standing problem of measuring the complete temporal (or, equivalently, spectral) intensity, and phase of arbitrary ultrashort pulses without prior assumptions about the pulse shape. It operates by spectrally resolving a signal field, $E_{sig}(t,\tau)$, generated in a nonlinear optical process by the pulse and, typically, its variably delayed replica(s). This provides a spectrogram—a two-dimensional data array vs. frequency $\omega$ and delay $\tau$—called the FROG trace, $I_{FROG}(\omega,\tau)$:

$$I_{FROG}(\omega,\tau) = \left| \int_{-\infty}^{+\infty} E_{sig}(t,\tau)\exp(-i\omega t)\, dt \right|^2 . \qquad (1)$$

FROG measures ultrashort pulses in many regions of spectrum (IR to XUV) and over a wide range of pulse complexities and temporal durations (femtoseconds to nanoseconds) [1], [2]. Direct inversion of the above expression for the pulse field (that is, the intensity and phase vs. time or frequency) is not possible, so an iterative phase-retrieval algorithm must be used to retrieve the full temporal information of the pulse.

FROG and other pulse-measurement techniques usually average over many pulses. So, the capabilities of various modern pulse-measurement techniques were recently studied for measurements in the presence of various types of instabilities in pulse trains that commonly arise in practice, such as partial mode-locking and unstable double-pulsing (which occurs when over-pumping the laser). These studies revealed that some pulse-measurement methods yield a much shorter pulse (the coherent component of the unstable train, usually called the coherent artifact) with little or no indication of instability. Since no pulse-stability meter exists, using a technique that measures only the coherent artifact is highly undesirable. The class of technique that performed best in these studies, even usually returning a pulse with the typical features of the fluctuating pulses for the given train of pulses, was FROG. Also, disagreement between the measured and retrieved FROG traces reflects the instability of the train [3]-[8].

Unfortunately, in addition to feedback on the pulse-train stability, a discrepancy between the measured and retrieved FROG traces could also occur if the pulse retrieval-algorithm stagnates, that is, does not converge to the correct pulse, for a given trace in the absence of instability. The performance of the best-known and most commonly used FROG algorithm, Generalized Projections (GP), has been studied previously for several FROG variations, and it was found that, except for the XFROG variant (which requires using a known reference pulse and so is not as useful as other versions of FROG that do not), stagnation occurs a good fraction of the time—more commonly as pulses increase in complexity [9]. Due to the possibility of stagnation, when disagreement between the traces occurs, the algorithm typically uses another (usually random) initial guess for the field. But, when discrepancies between the measured and retrieved traces persist, it can be difficult to know when to give up and conclude that the discrepancies are due to pulse-train instability and not stagnation. This is particularly important for amplified pulses, which are more prone to instability than unamplified, high-rep-rate oscillators. As a result, an active area of ongoing research is the development of a more robust reconstruction algorithm for FROG [10], [11]. Indeed, it can be realistically argued that the most important unsolved problem in the field of pulse measurement today is the development of a 100% reliable algorithm for the various variants of FROG.

So, here, we solve the problem of stagnation of FROG's pulse-retrieval algorithm by introducing what we call the Retrieved-Amplitude N-grid Algorithmic (RANA) approach. The RANA approach involves using the standard GP algorithm (or any other FROG algorithm the user desires), but first obtaining a *vastly improved initial guess*. Specifically, instead of using random noise, the autocorrelation, or a flat phase, fixed-length Gaussian pulse, as is usually done, we *directly* retrieve the *precise pulse spectrum* from the measured FROG trace. This is accomplished entirely from the trace marginals—integrals of the trace over delay and frequency. While this procedure is in fact quite simple, it has not been realized previously. Then we generate a set of pulses, all with the correct spectrum, but each with random noise for the spectral phase. We then quickly run them using the GP algorithm on coarser, smaller grids generated from the full trace, removing the most poorly performing pulses and keeping only the best-performing ones [12], [13]. This rapidly yields an excellent initial guess for the *spectral phase* also. This is all done before the full trace is considered in the algorithm. As a result, only a few iterations using the entire trace (the slowest step in any algorithm) are required before the correct pulse is retrieved. We use the RANA approach for both second-harmonic generation (SHG) FROG [14] (which we consider elsewhere) and also polarization-gate (PG) FROG and transient-grating (TG) FROG, the latter two of which we describe here. In each of these versions of FROG, we observe *zero* stagnations in over 20,000 noise-corrupted traces corresponding to simple and extremely complex pulses with time-bandwidth products (TBPs) as large as 100.

## II. Polarization-Gate FROG

Polarization-gate (PG) FROG [1] uses a third-order nonlinear-optical effect, in which the variably delayed gate pulse induces birefringence in a medium by the optical Kerr effect, and hence results in causing the medium to act as a wave-plate for the probe pulse passing through it, as shown in Fig. 1. This process can be described mathematically as $E_{sig}(t,\tau) = E(t)\ |E(t\text{-}\tau)|^2$. The nonlinear-optical process is automatically phase-matched, so there is no limitation on the bandwidth of the pulse to be measured. In addition, the conversion efficiency is essentially independent of wavelength [15]. As a result, PG FROG is used for measurement of tunable, broadband, or UV pulses, for which SHG crystals are not available. Unlike SHG FROG, PG FROG also has the advantage that its traces are intuitive and mirror the instantaneous frequency vs. time, so that, for example, positive and negative chirp can be distinguished directly from the traces [16], [17]. Also, unlike SHG FROG, it also has no ambiguity in the direction of time. As a result, PG FROG is often used in pulse-shaping applications [18]. For these reasons, PG FROG is a very popular pulse-measurement technique, especially for amplified pulses, which have the intensity needed for such a third-order effect and are more likely to be unstable and/or complex.

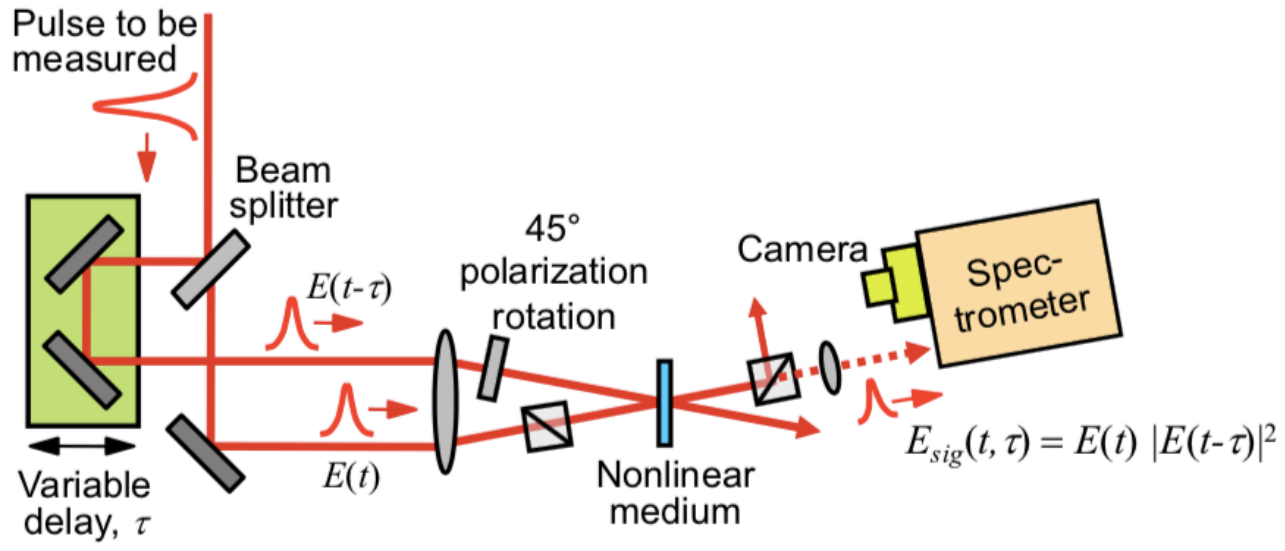

Fig. 1. Polarization-gate FROG schematic. Reprinted with permission from [19].

In addition to the PG geometry, two of the three possible FROG beam geometries involving three beams, called transient-grating (TG) FROG (Fig. 2), yield traces that are mathematically equivalent to those of PG FROG [20]. TG FROG is now commonly used in the measurement of high intensity laser pulses [21]-[23], few-cycle pulses (as there are no elements in the setup that yield significant dispersion, such as polarizers) [24]-[27], and deep UV pulses [28], [29]. Moreover, the lack of background caused by polarizer leakage and the use of a diagonal element of the nonlinear medium's $\chi^{(3)}$ tensor make it a more sensitive technique than PG FROG.

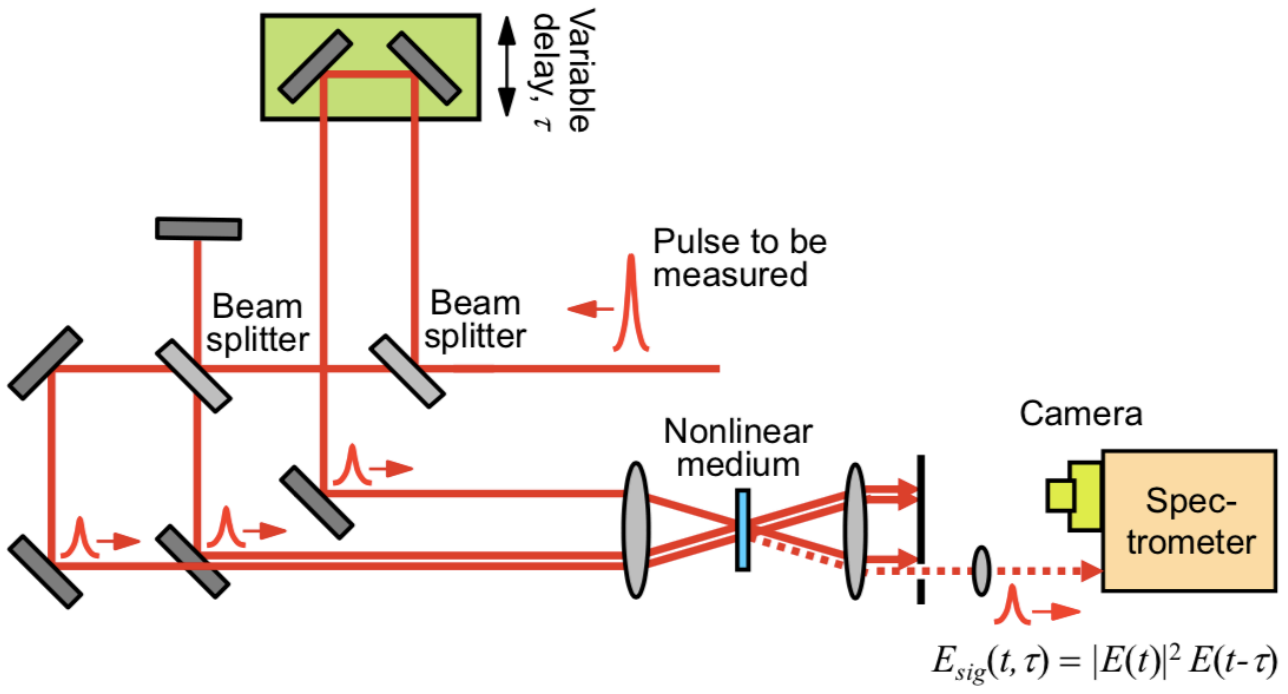

Fig. 2. Transient-grating FROG schematic.

Due to their importance, we demonstrate here the RANA approach for the PG and TG geometries and show that it provides 100% convergence for over 20,000 pulses of varying complexities and in the presence of noise. We also find that it is faster than the standard GP algorithm, especially for complex pulses.

## III. The RANA Approach

The RANA approach involves first *retrieving the precise spectral intensity of the pulse directly from the FROG trace marginals*. A vastly improved initial guess for the spectral *phase* is also obtained by quickly running pulses with the correct spectrum and random spectral phases through smaller, coarser arrays derived from the measured trace and choosing the resulting pulses that best approximate these smaller arrays. Using such vastly improved initial guesses in both the spectral intensity and phase, the standard GP algorithm (or any other FROG algorithm) then only requires a few iterations using the complete trace to retrieve the correct pulse in all cases [14].

We first show how to obtain the spectrum from the PG FROG (or TG FROG) trace. The frequency marginal in PG FROG can be written as [30]:

$$\begin{aligned} M^{PG}(\omega) &= \int_{-\infty}^{+\infty} I_{FROG}^{PG}(\omega,\tau)\,d\tau \\ &= \int_{-\infty}^{+\infty} \left| \int_{-\infty}^{+\infty} E(t)\left|E(t-\tau)\right|^2 \exp(-i\omega t)\,dt \right|^2 d\tau \\ &= S(\omega) * \mathscr{F}\left\{A^{(2)}(\tau)\right\}. \end{aligned} \quad (2)$$

where $S(\omega)$ is the spectrum, and $A^{(2)}(\tau)$ is the second-order intensity autocorrelation (AC) of the pulse. Based on (2), if the second-order AC of the pulse is available, the spectrum can be retrieved using a quickly converging iterative deconvolution algorithm or simply the convolution theorem:

$$S(\omega) = \mathscr{F}\left\{ \frac{\mathscr{F}^{-1}\left\{M^{PG}(\omega)\right\}}{A^{(2)}(\tau)} \right\}, \quad (3)$$

when $A^{(2)}(\tau) \neq 0$. Of course, $A^{(2)}(\tau)$ does approach zero in its wings. But so does $\mathscr{F}^{-1}\{M^{PG}(\omega)\}$. Usually the values for both marginals are merely very small and not exactly zero. But for the case $A^{(2)}(\tau) \approx 0$, the neighboring values of the AC for that delay are used in the division. However, another issue arises, which is more challenging: for PG and TG FROG, it is not actually possible to obtain $A^{(2)}(\tau)$ rigorously - from the PG trace because (unlike SHG FROG) the PG FROG delay marginal is equivalent to the *third-order* AC, $A^{(3)}(\tau)$.

Fortunately, $A^{(3)}(\tau)$ can be modified to approximate $A^{(2)}(\tau)$ for our purposes. Because $A^{(2)}(\tau)$ is always symmetrical with respect to delay, we begin by symmetrizing the delay marginal, $M^{PG}(\tau) = A^{(3)}(\tau)$, by computing the average value of $A^{(3)}(\tau)$ and $A^{(3)}(-\tau)$. Because $A^{(2)}(\tau)$ is generally wider than $A^{(3)}(\tau)$, we also raise the modified third-order AC, $A_s^{(3)}(\tau)$ to a power, $p$, smaller than one:

$$S_{retrieved}(\omega) = \mathscr{F}\left\{ \frac{\mathscr{F}^{-1}\left\{M^{PG}(\omega)\right\}}{\left(A_s^{(3)}(\tau)\right)^p} \right\}. \quad (4)$$

To determine the optimal value of $p$, we tested a range of values between 0.6 and 0.8, and determined the rms difference between the second-order AC and the scaled and modified third-order AC for a set of sample pulses using the expression below:

$$\sqrt{\frac{1}{N}\sum_{i=1}^{N}\left|A^{(2)}(\tau_i) - \mu\left(A_s^{(3)}(\tau_i)\right)^p\right|^2} \quad (5)$$

where $\mu$ is a constant that minimizes this difference, and $A^{(2)}(\tau)$ has the peak intensity of 1. By determining the best value for $p$, we retrieve the spectral intensity of the pulse from the trace marginal remarkably reliably—even in the presence of significant noise in the trace. Our results using this approach are described in the next section.

Next, in order to also obtain a better guess for the spectral phase than the usual random noise or other functions, we generate a set of smaller and coarser grids from the full $N \times N$ trace with dimensions of $N/2 \times N/2$ and $N/4 \times N/4$ and begin the usual GP algorithm on the smallest trace using multiple initial guesses (all using the precise spectrum obtained as described above, but random spectral phase). Also, in transitioning from a smaller trace to the next larger one, we re-apply the above directly retrieved spectrum to the retrieved field if doing so improves the agreement between the resulting pulse's third-order AC and the delay marginal. We also remove the poorest pulses and keep only the best ones based on the $G$ error values (the minimum rms difference between the measured and retrieved traces) [1].

Finally, the four best spectral phases obtained from the $N/2 \times N/2$ trace are then used in combination with the correct spectrum as initial guesses for the entire $N \times N$ array, for which only a few iterations are required. And, although convergence always occurred for the first pulse tried, to ensure absolute reliability, we retained four pulses for these final few iterations. The multi-grid part of the RANA approach is depicted in Fig. 3.

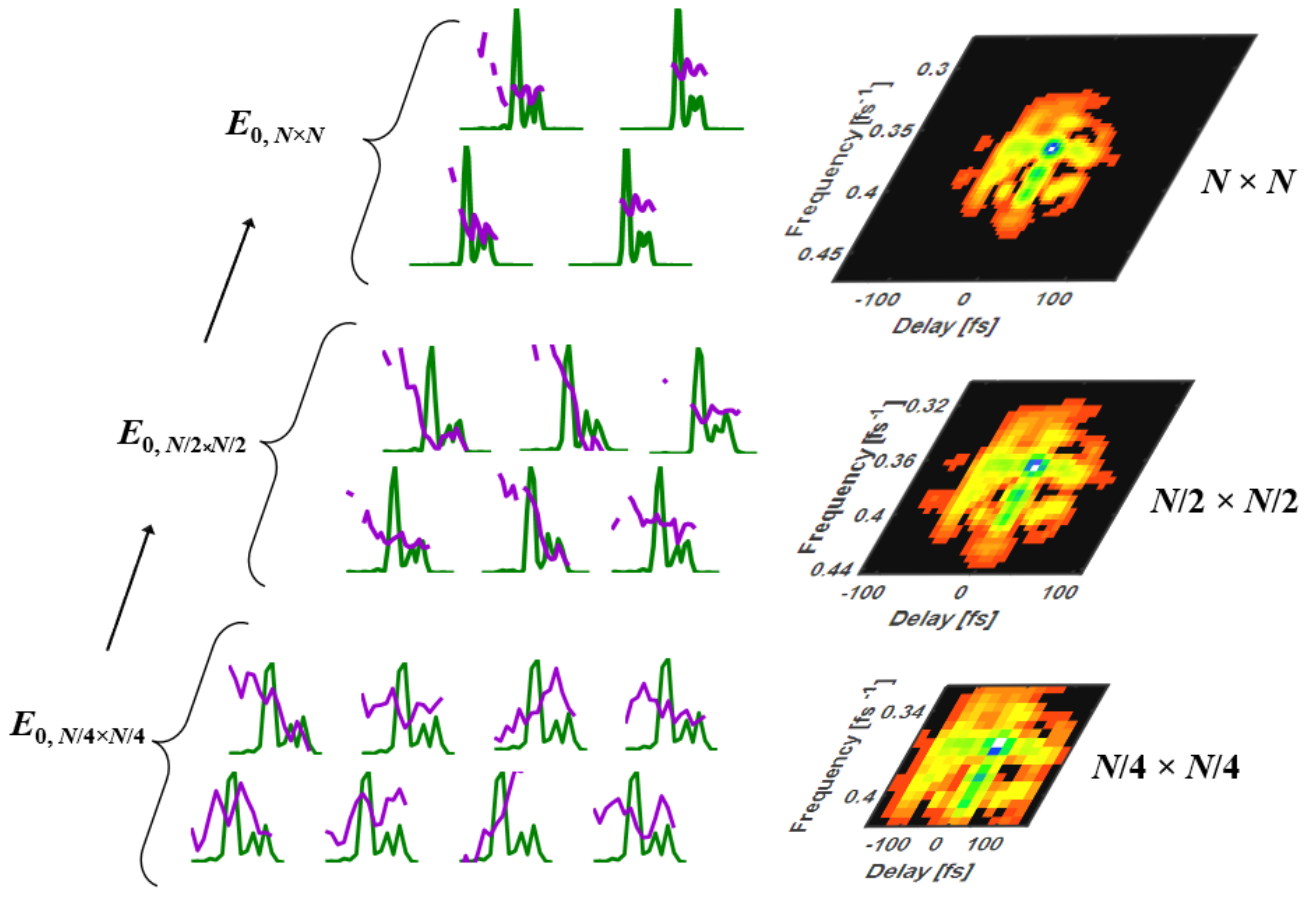

Fig. 3. Graphical representation of the multi-grid component of the RANA approach. $E_0$ corresponds to the set of initial guesses used on each array.

It should be noted that, due to the trivial ambiguity in the arrival time of the retrieved field in the temporal domain, the generation of the next initial guess from the retrieved results (adding zeros to the sides of the retrieved field to extend the temporal/spectral range and interpolating to the proper size) should be done in the spectral domain rather than the temporal domain. The retrieved spectra were also multiplied by a super-Gaussian of order six, for which the intensity would go to nearly zero for ~10% of the points in the wings before performing pulse retrieval.

This process also benefits from parallel processing features provided by MATLAB (a four-core processor was used for this work, hence the use of four initial guesses for iterations using the full trace). Specifically, the RANA approach, which uses multiple pulses simultaneously, more naturally parallelizes than the usual GP approach. Of course, in this work, we used the four-core parallel processing available in our computer for both approaches, although the RANA approach benefitted from it more.

We note here that the RANA approach for SHG FROG [14], which is described in more detail in a separate publication, varies slightly from that for PG FROG. Unlike (2), the frequency marginal for SHG FROG is simply the *autoconvolution* of the spectrum. The convolution theorem then provides that a simple square-root operation must be performed, yielding two possible roots at every point. In order to choose the correct root, it is simply necessary to note that the Paley-Wiener Theorem provides that the inverse Fourier-transform of the spectrum is continuous, and as are all of its derivatives. The remaining steps in the RANA approach for SHG FROG are analogous to the approach described above. It performs as well for SHG FROG as it does for PG and TG FROG, that is, no stagnations for over 20,000 noise-corrupted traces of pulses with TBPs as high as 100.

To assess the RANA-approach performance for PG/TG FROG, we simulated a set of random pulses with rms *TBPs* of 2.5, 5, 10, 20, 40, 80 and 100, where 0.5 is the *TBP* for a Fourier-transform-limited Gaussian. Next, the simulated traces were contaminated with 1% of multiplicative noise plus 1% additive noise.

In view of the known uniqueness of FROG (except for a few known trivial ambiguities), to determine whether the algorithm converges to the correct field we used the $G$ and $G'$ [31] errors (rms errors between the measured and retrieved traces) as the measures for the convergence of the retrieval algorithm. This eliminates having to deal with the trivial ambiguities. The corresponding values of $G$ and $G'$ for a reliable result for a retrieval that yields agreement between the simulated pulse and the retrieved one are given in Table 1. If either of the cutoff values for $G$ or $G'$ errors were obtained, the algorithm was considered to have converged. This approach was confirmed by visual inspection of the retrieved pulses with the highest $G$-errors and confirming that they agreed within experimental error (that is, the noise values) with the actual pulses.

TABLE I
PARAMETERS—THE NUMBER OF INITIAL GUESSES (IGS) AND ITERATIONS FOR EACH ARRAY SIZE—USED IN THE IMPLEMENTATION OF RANA APPROACH TO THE PG FROG PULSE-RETRIEVAL ALGORITHM. NOTE THAT, DESPITE THE NOISE ADDED, OUR CRITERIA ARE MORE STRINGENT THAN THE USUAL $G < 1\%$ RULE OF THUMB.

| | | ***N*/4 × *N*/4** | | ***N*/2 × *N*/2** | | ***N* × *N*** | | |
|---|---|---|---|---|---|---|---|---|
| **Pulse *TBP*** | **Array size, *N*** | **# of IGs** | **# of iterations** | **# of IGs** | **# of iterations** | **# of IGs** | **Maximum *G* /*G*' error** | **# of sample pulses** |
| 2.5 | 64 | 12 | 20 | 8 | 20 | 4 | 0.0090/0.2 | 5000 |
| 5 | 128 | 12 | 25 | 8 | 20 | 4 | 0.0080/0.2 | 5000 |
| 10 | 256 | 20 | 25 | 12 | 25 | 4 | 0.0070/0.2 | 5000 |
| 20 | 512 | 24 | 30 | 16 | 25 | 4 | 0.0065/0.2 | 5000 |
| 40 | 1024 | 28 | 35 | 16 | 30 | 4 | 0.0045/0.2 | 2500 |
| 80 | 2048 | 36 | 40 | 24 | 35 | 4 | 0.0035/0.2 | 250 |
| 100 | 4096 | 48 | 40 | 28 | 35 | 4 | 0.0020/0.2 | 100 |

## IV. RESULTS

In order to determine the value for the power, $p$, we used a set of sample pulses with different complexities. The average of rms differences, as defined by (5), were determined between the second-order and modified third-order AC vs. $p$ (as shown in Fig. 4). Based on these results, we chose $p$ = 0.73. Fig. 5 shows the effect of this modification on the delay marginal and the resulting retrieved spectrum using (4).

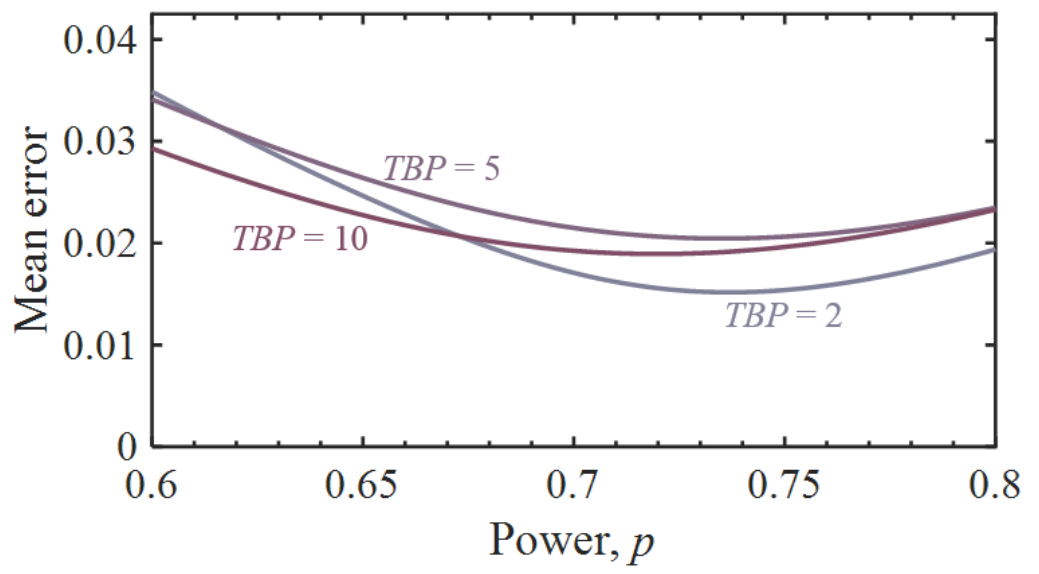

Fig. 4. The average rms error between, the second-order intensity autocorrelation, $A^{(2)}(\tau)$, and the modified third-order intensity autocorrelation, $[A_s^{(3)}(\tau)]^p$, for a set of simulated pulses with *TBP*s of 2, 5, and 10. Based on these plots, the value of $p$ = 0.73 was chosen for approximating delay marginal of PG trace as $A^{(2)}(\tau)$.

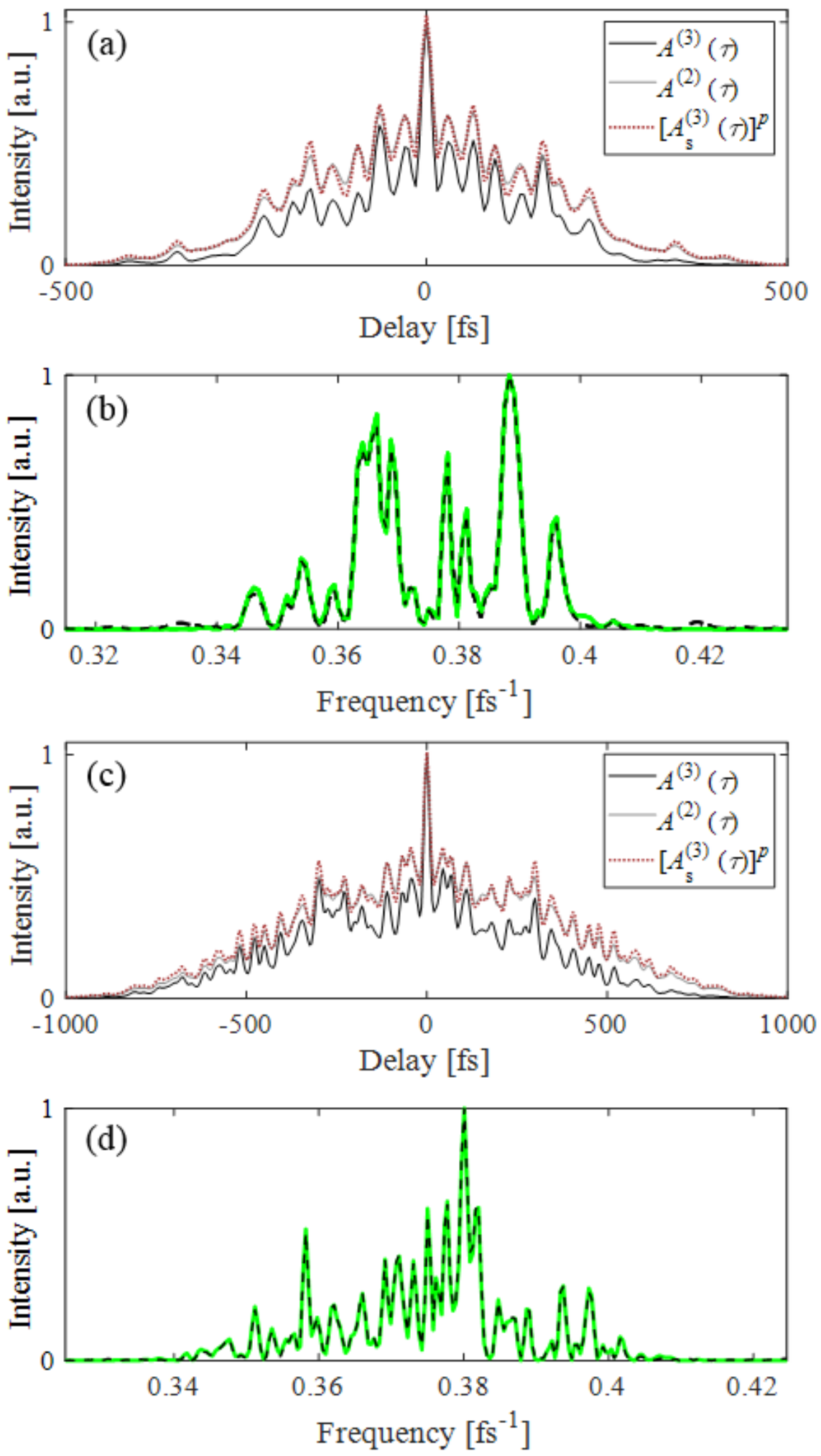

Fig. 5. (a, b) Normalized delay marginals of two different PG FROG traces (black), the second-order ACs (gray), and the scaled symmetrized delay marginals raised to a power of 0.73 (dotted brown) (for *TBP* = 10 and 20, respectively). (c, d) The simulated and retrieved spectra from the frequency marginals of the PG traces using the modified delay marginals shown in (a) and (b) and using (4), respectively. Note that the modified delay marginals very closely approximate the second-order ACs, and the retrieved spectra are also determined quite precisely. This was true for all pulses considered.

Also, this procedure is independent of the temporal (spectral) sampling values as shown in Fig. 6. This is because the division happens in the same domain (time), and the inverse Fourier-transform of $M(\omega)$ has smaller temporal width than the AC.

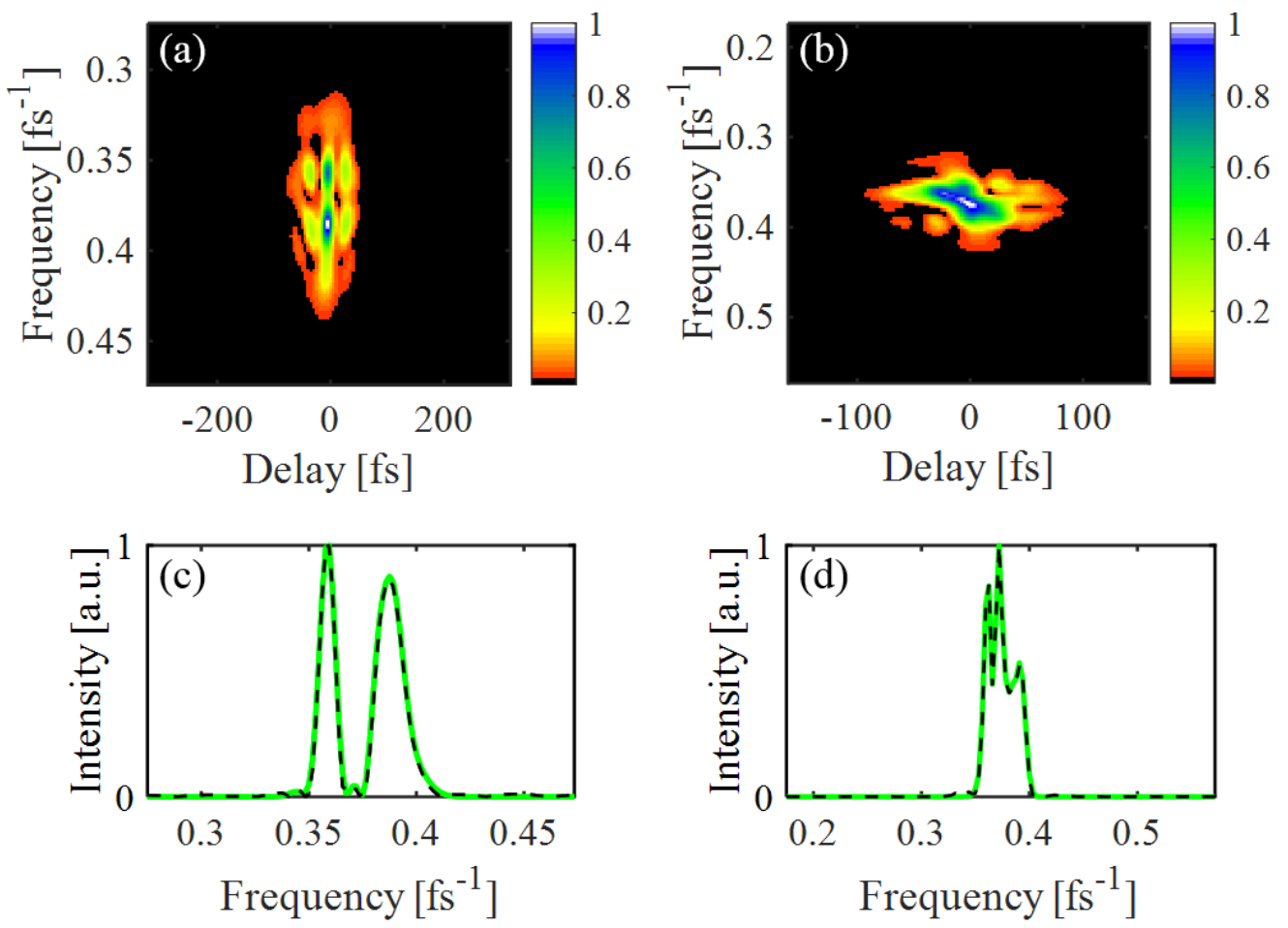

Fig. 6. Spectra obtained from the marginals of PG traces with uneven temporal and spectral distributions (a, b). The simulated and retrieved spectra are plotted in light and dashed black, respectively, and agree very well in both cases (c, d).

Fig. 7 shows the four spectra, corresponding to the quartiles of rms errors and the largest value of rms error obtained using (5), that are retrieved directly from noisy PG FROG traces for two different *TBP*s. Note that the retrieved spectra are very close to the actual spectra, despite the approximations used and the noise added to the traces. The directly retrieved traces, even in the worse cases, would generally even be considered acceptable for the final resulting spectrum, although, in the RANA approach, they are only as the initial guess and so are improved further by the algorithm.

Fig. 8 demonstrates the performance of the RANA approach in retrieving the spectral phase.

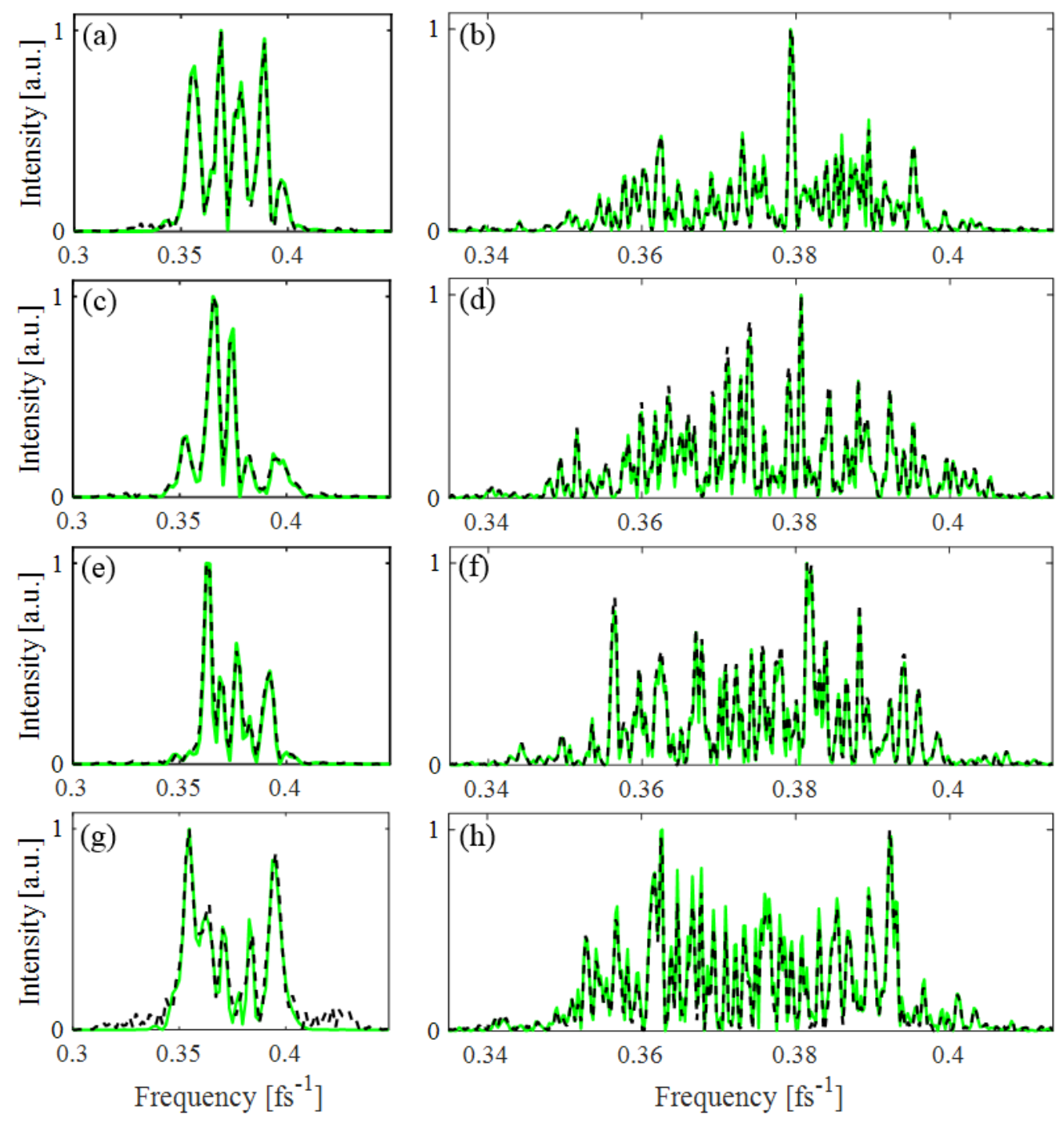

Fig. 7. The performance of direct spectral retrieval from the marginals for two pulse sets with *TBP* = 5 and 40 (first and second columns, respectively). (a, b) correspond to the best results (lowest quartile rms errors) between the simulated spectra (solid light green) and the retrieved ones (dashed black). (c, d) and (e, f) correspond to results for which rms error is the middle quartile (middle 50%) and upper quartiles, respectively. Note that, even for the worst cases (g and h), agreement between actual spectra and that retrieved directly from the marginals is excellent. Of course, these spectra are only used as initial guesses, so further improvement occurs in the eventual iterative process using the full measured FROG traces.

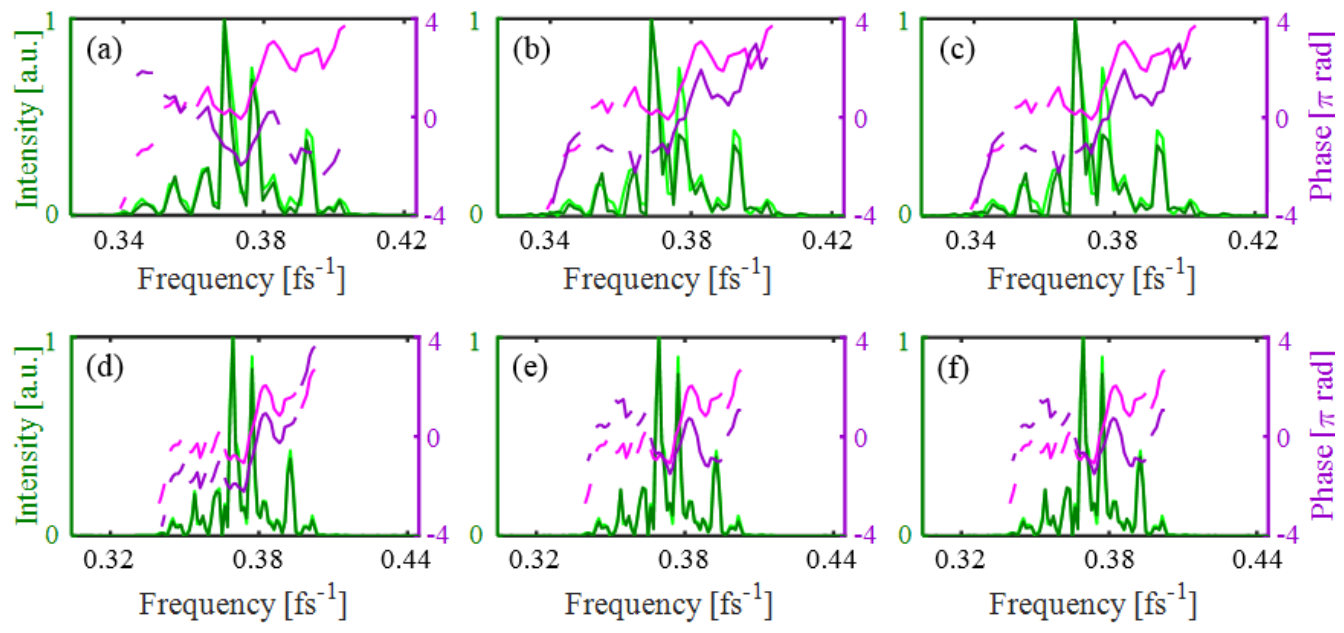

Fig. 8. The spectral intensity (dark green) and phase (purple) retrieved after the *N*/4 × *N*/4 array and then the *N*/2 × *N*/2 array are shown in top and bottom rows, respectively, for a 256 × 256 trace with *TBP* = 10. The binned simulated spectral intensities and phases are plotted in light green and magenta, respectively. (a, d), (b, e), and (c, f) correspond to lowest (best), intermediate, and highest *G* error (worst) retrievals, respectively. The phases are separated in order to better compare them and also because the absolute (zero-order) phase is arbitrary and not measured by FROG. Note that even the worst of these pulses are very close to the actual pulses and so provide excellent initial guesses for the iterations on the complete trace.

Fig. 9 shows a typical pulse used for testing the RANA approach for PG FROG, and the corresponding spectrum that is retrieved directly from its trace and used as the spectral intensity for the initial guesses.

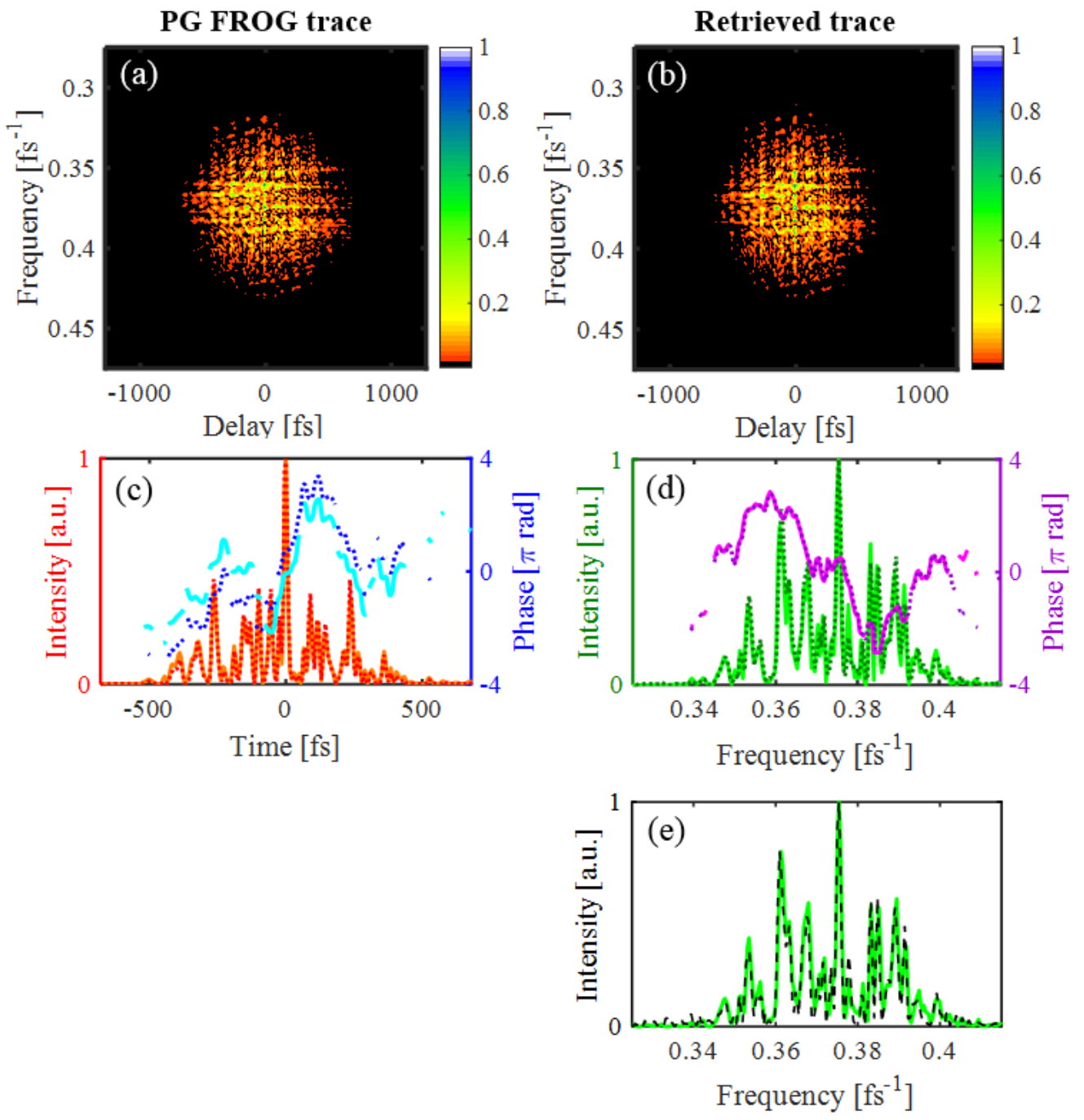

Fig. 9. (a) Simulated PG FROG trace with 1% multiplicative and 1% additive noise (after noise removal using a low-pass filter and background subtraction, as is usually done in FROG measurements), and (b) the retrieved trace. (c, d) The reconstructed temporal/spectral intensity and phase are shown in red/green and blue/purple. The simulated temporal/spectral intensity and phase are shown in orange/light-green and cyan/magenta. The *G* error for retrieval of this 512 × 512 trace is 0.0062 and *G′* error = 0.240, both indicating excellent convergence. (e) The directly retrieved spectral intensity from the marginals is shown as the dashed black curve, and the spectrum of the actual pulse is the light green curve.

Fewer pulses with the highest complexities were considered due to the lengthy computer runs involved. Of course, use of a faster programming language than the notoriously slow MATLAB that we used will yield correspondingly faster convergence with the same reliability. Our results for the convergence time should therefore only be use for comparison; in practice, with faster code, much faster convergence times will occur. As mentioned, the *G* and *G′* errors were used as the measures for convergence of the retrieval algorithm, and they were never higher than the maximum acceptable value for any of the cases that we tried. In other words, the RANA approach converged for *all* of the more than 20,000 sample pulses with different, and even very high, complexities.

TABLE 2
COMPARISON OF PERFORMANCE OF STANDARD GP ALGORITHM AND THE RANA APPROACH.

| **Pulse *TBP*** | **Array size, *N*** | **Average retrieval time for converging initial guess** | **GP algorithm percentage of convergence on first initial guess** | **# of sample pulses** | **Average retrieval time** | **RANA approach percentage of convergence on first initial guess** |
|---|---|---|---|---|---|---|
| 2.5 | 64 | 0.164 s | 90.7% | 5000 | 0.191s | 100% |
| 5 | 128 | 0.851 s | 83.4% | 5000 | 0.354 s | 100% |
| 10 | 256 | 2.91 s | 69.8% | 5000 | 1.42 s | 100% |
| 20 | 512 | 12.9 s | 65.8% | 5000 | 7.15 s | 100% |
| 40 | 1024 | 75.1 s | 68.0% | 2500 | 38.1 s | 100% |
| 80 | 2048 | 12.2 min | 57.1% | 250 | 5.28 min | 100% |
| 100 | 4096 | 64.5 min | 40.0% | 100 | 24.0 min | 100% |

Due to the differences in the implementations of the GP algorithm for SHG and PG FROG, the time per iteration is longer for the PG algorithm, and so the retrieval times are longer than those in [14].

## V. DISCUSSION

It is interesting to assess the relative contributions of the two innovations of the RANA approach used here for PG FROG: 1) the directly-retrieved-spectrum initial guess and 2) the multi-grid/multi-initial-guess strategies. The benefit of using the spectrum directly retrieved from the marginals of the trace can be observed in the convergence percentage of the GP algorithm when it is used without using multiple initial guesses or smaller grids. We found that, for pulse sets with *TBP's* of 20 and 40, the convergence of the simple algorithm increases from 65.8% to 81.2%, and from 68.9% to 77.6%, respectively. Use of the approximate spectrum also improved the average retrieval time, which dropped from 12.9s to 8.32s and from 75.1s to 63.9s, respectively. Thus, the multi-grid and multi-initial guess components of the RANA approach also play a key role, not only in the convergence time, but also in the convergence percentage of the algorithm. The RANA approach actually spent about 90% of its time on the smaller grids, and, by doing so, it provides a excellent initial guess for the spectral phase on the full grid.

## VI. CONCLUSIONS

In order to address the need for a highly reliable (and fast) phase-retrieval algorithm for FROG, we have developed a new and powerful algorithmic approach, which we call the RANA approach and here describe its operation for two of the most popular variations of FROG: PG and TG. It is especially important for the characterization of amplified pulses, broadband pulses, and/or pulses in the UV region of spectrum and also in the presence of pulse-shape instability. It retrieves the spectrum of the pulse directly from the marginals of the trace reliably even in the presence of additive and multiplicative noise. Smaller, coarser trace grids then allow us to obtain a more accurate initial spectral phase, as well. Thus, we have obtained a significantly improved guess for the entire pulse before the required use of the entire trace, which is the slowest step in any FROG algorithm. As a result, the RANA approach proved extremely robust—and also faster, as only a few (typically four) iterations using the entire trace proved necessary.

We tested the RANA approach on more than 20,000 pulses and their corresponding noise-contaminated PG/TG FROG traces, and we achieved convergence for all the pulses, as well as shorter retrieval times for complex pulses. As the RANA approach is effectively a technique for generating a vastly improved initial guess, it can also be used with any FROG algorithm, not just the usual GP algorithm, so it will benefit from future FROG algorithms that could be faster than the standard GP algorithm that we used. We conclude that the RANA approach provides a much-needed, perfectly reliable algorithmic approach for PG, TG, and SHG FROG that will benefit experimenters in a wide range of applications of ultrashort pulses.

## FUNDING

This work was supported by the National Science Foundation under Grant ECCS-1307817 and the Georgia Research Alliance. Rick Trebino owns a company that sells pulse-measurement devices.

## REFERENCES

[1] R. Trebino, *Frequency-Resolved Optical Gating: The Measurement of Ultrashort Laser Pulses*. MA: Kluwer Academic Publishers, 2002.
[2] P. Bowlan and R. Trebino, "Complete single-shot measurement of arbitrary nanosecond laser pulses in time," *Opt. Express,* vol. 19, no. 2, pp. 1367-77, 2011.
[3] J. Ratner, G. Steinmeyer, T. C. Wong, R. Bartels, and R. Trebino, "Coherent artifact in modern pulse measurements," *Opt. Lett.,* vol. 37, no. 14, pp. 2874-2876, 2012.
[4] M. Rhodes, G. Steinmeyer, J. Ratner, and R. Trebino, "Pulse-shape instabilities and their measurement," *Laser Photon. Rev.* , vol. 7, no. 4, pp. 557-565, 2013.
[5] M. Rhodes, M. Mukhopadhyay, J. Birge, and R. Trebino, "Coherent artifact study of two-dimensional spectral shearing interferometry," *J. Opt. Soc. of Am. B,* vol. 32, no. 9, pp. 1881-1888, 2015.
[6] M. Rhodes and R. Trebino, "Coherent Artifact Study of Multiphoton Intrapulse Interference Phase Scan," in *CLEO: 2015*, San Jose, California, 2015, p. JTu5A.15: Optical Society of America.

[7] M. Rhodes, Z. Guang, and R. Trebino, "Unstable and Multiple Pulsing Can Be Invisible to Ultrashort Pulse Measurement Techniques," *App. Sci.,* vol. 7, no. 1, p. 40, 2017.

[8] R. Jafari, E. Esmerando, and R. Trebino, "Linear-chirp instability in ultrashort-laser-pulse measurement," *Submitted for publication*, 2018.

[9] L. Xu, E. Zeek, and R. Trebino, "Simulations of frequency-resolved optical gating for measuring very complex pulses," *J. Opt. Soc. of Am. B,* vol. 25, no. 6, pp. A70-A80, 2008.

[10] P. Sidorenko, O. Lahav, Z. Avnat, and O. Cohen, "Ptychographic reconstruction algorithm for frequency-resolved optical gating: super-resolution and supreme robustness," *Optica,* vol. 3, no. 12, pp. 1320-1330, 2016.

[11] R. Aboushelbaya *et al.*, "Single-shot frequency-resolved optical gating for retrieving the pulse shape of high energy picosecond pulses," *Rev. Sci. Instrum.,* vol. 89, no. 10, p. 103509, 2018.

[12] C. W. Siders, J. L. W. Siders, F. G. Omenetto, and A. J. Taylor, "Multipulse interferometric frequency-resolved optical gating," *IEEE J. of Quantum Electron.,* vol. 35, no. 4, pp. 432-440, 1999.

[13] R. Jafari and R. Trebino, "High-speed “multi-grid” pulse-retrieval algorithm for frequency-resolved optical gating," *Opt. Express,* vol. 26, no. 3, pp. 2643-2649, 2018.

[14] R. Jafari, T. Jones, and R. Trebino, "100% reliable algorithm for second-harmonic-generation frequency-resolved optical *gating (Accpted for publication)*," *Opt. Express*, to be published.

[15] H. Kano and H.-o. Hamaguchi, "Characterization of a supercontinuum generated from a photonic crystal fiber and its application to coherent Raman spectroscopy," *Opt. Lett.,* vol. 28, no. 23, pp. 2360-2362, 2003

[16] D. J. Kane and R. Trebino, "Characterization of arbitrary femtosecond pulses using frequency resolved optical gating," *IEEE J. Quantum Electron.,* vol. 29, no. 2, pp. 571-579, 1993.

[17] B. C. Walker *et al.*, "A 50-EW/cm2 Ti:sapphire laser system for studying relativistic light-matter interactions," *Opt. Express,* vol. 5, no. 10, pp. 196-202, 1999.

[18] K. Ohno, T. Tanabe, and F. Kannari, "Adaptive pulse shaping of phase and amplitude of an amplified femtosecond pulse laser by direct reference to frequency-resolved optical gating traces," *Journal of the Opt. Soc. of Am. B,* vol. 19, no. 11, pp. 2781-2790, 2002.

[19] R. Trebino. (11/2018). *Ultrafast Optics Group*. Available: http://frog.gatech.edu/

[20] D. Lee, S. Akturk, P. Gabolde, and R. Trebino, "Experimentally simple, extremely broadband transient-grating frequency-resolved-optical-gating arrangement," *Opt. Express,* vol. 15, no. 2, pp. 760-766, 2007.

[21] S. Backus, C. G. Durfee, G. Mourou, H. C. Kapteyn, and M. M. Murnane, "0.2-TW laser system at 1 kHz," *Opt. Lett.,* vol. 22, no. 16, pp. 1256-1258, 1997.

[22] S. Backus, C. G. D. III, M. M. Murnane, and H. C. Kapteyn, "High power ultrafast lasers," *Rev. Sci. Instrum.,* vol. 69, no. 3, pp. 1207-1223, 1998.

[23] C. G. Durfee, S. Backus, M. M. Murnane, and H. C. Kapteyn, "Design and implementation of a TW-class high-average power laser system," *IEEE J. of Sel. Top. in Quantum Electron.,* vol. 4, no. 2, pp. 395-406, 1998.

[24] M. Li, J. P. Nibarger, C. Guo, and G. N. Gibson, "Dispersion-free transient-grating frequency-resolved optical gating," *Appl. Opt.,* vol. 38, no. 24, pp. 5250-5253, 1999.

[25] T. Fuji, T. Horio, and T. Suzuki, "Generation of 12 fs deep-ultraviolet pulses by four-wave mixing through filamentation in neon gas," *Opt. Lett.s,* vol. 32, no. 17, pp. 2481-2483, 2007.

[26] J. N. Sweetser, D. N. Fittinghoff, and R. Trebino, "Transient-grating frequency-resolved optical gating," *Opt. Lett.,* vol. 22, no. 8, pp. 519-521, 1997.

[27] A. S. Pirozhkov *et al.*, "Transient-grating FROG for measurement of sub-10-fs to few-ps amplified pulses," in *Advanced Solid-State Photonics*, Nara, 2008, p. MC8: Optical Society of America.

[28] T. Nagy and P. Simon, "Single-shot TG FROG for the characterization of ultrashort DUV pulses," *Opt. Express,* vol. 17, no. 10, pp. 8144-8151, 2009.

[29] F. Lindau *et al.*, "Pulse Shaping at the MAX IV Photoelectron Gun Laser," *Proc. IPAC’17,* pp. 1541-1543, 2017.

[30] K. W. DeLong, R. Trebino, and D. J. Kane, "Comparison of ultrashort-pulse frequency-resolved-optical-gating traces for three common beam geometries," *J. Opt. Soc. of Am. B,* vol. 11, no. 9, pp. 1595-1608, 1994.

[31] R. P. Scott *et al.*, "High-fidelity line-by-line optical waveform generation and complete characterization using FROG," *Opt. Express,* vol. 15, no. 16, pp. 9977-9988, 2007.